\documentclass{eas}
\usepackage{graphicx}
\usepackage[T1]{fontenc}
\usepackage[latin2]{inputenc}
\begin{document}

\title{Is there a metallicity enhancement in planet hosting red giants?} 
\author{Pawe\l{} Zieli\'nski}\address{Toru\'n Centre for Astronomy, Nicolaus Copernicus University, ul. Gagarina 11, 87-100 Torun, Poland}
\author{Andrzej Niedzielski}\sameaddress{1}\secondaddress{Department of Astronomy and Astrophysics, Pennsylvania State University, 525 Davey Laboratory, University Park, PA 16802}
\author{Monika Adam\'ow}\sameaddress{1}
\author{Aleksander Wolszczan}\sameaddress{2,1}
\runningtitle{Is there a metallicity enhancement in PH red giants?}
\begin{abstract}
The Penn State/Toru\'n Centre for Astronomy Search for Planets Around Evolved Stars is a high-precision radial velocity (RV) survey aiming at planets detection around giant stars. It is based on observations obtained with the 9.2 m Hobby-Eberly Telescope. As proper interpretation of high precision RV data for red giants requires complete spectral analysis of targets we perform spectral modeling of all stars included in the survey. Typically, rotation velocities and metallicities are determined in addition to stellar luminosities and temperatures what allows us to estimate stellar ages and masses. Here we present preliminary results of metallicity studies in our sample. We search a metallicity dependence similar to that for dwarfs by comparing our results for a sample of 22 giants earlier than K5 showing significant RV variations with a control sample of 58 relatively RV-stable stars.
\end{abstract}
\maketitle
\section{Introduction}
Over 300 planet-hosting stars are known today, approximately 20 of them being evolved stars that have left the Main Sequence already. As the population of hosts of extrasolar planets grows fast it has become very interesting to look for statistical properties of these objects  and possible correlations between fundamental stellar parameters and planet occurrence to understand better the planetary formation scenario.\\
One of the most intriguing features of planet hosting stars,  the metallicity  enhancement for main-sequence dwarfs was found and described by Gonzalez (\cite{G97}) and Santos \etal\ (\cite{S01}, \cite{S04}). In the case of FGK-type dwarfs the spectroscopic studies by Fisher \& Valenti (\cite{F05}) have demonstrated that extrasolar planets tend to be discovered more often around stars revealing higher metallicity. 

Giants hosting exoplanets don't seems to be metal-rich objects (Pasquini \etal\ \cite{P07}; Hekker \etal\ \cite{H07}, \cite{H08}) but due to small statistics these results remain uncertain still. In this paper we attempt to verify the planet occurrence vs. host metallicity relation for giants using a sample of 80 late-type giants from our survey (Fig.~1).

\begin{figure}[htb]
\center
\includegraphics[width=7cm, angle=270]{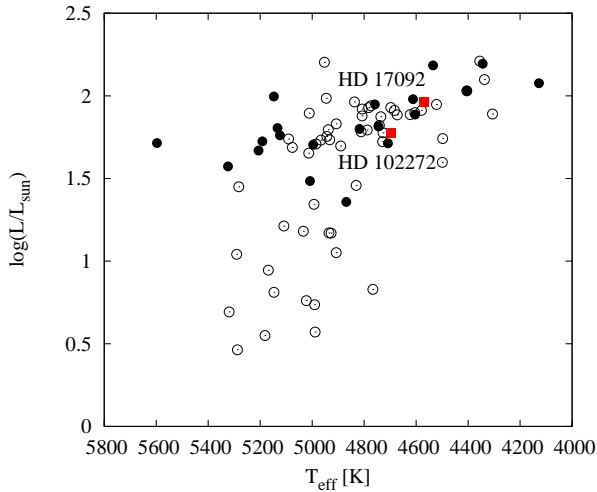}
\caption{HR diagram for our sample of 22 RV-variable stars (filled symbols) and 58 RV-stable stars (open symbols). Quadratic points present the two planet-hosting giants (HD~17092 and HD~102272) from this survey.}
\end{figure}

\section{Observations and data reduction}
The observational material used in this work are high quality, high-resolution optical spectra of late-type stars observed within our survey. Most of them are unstudied before objects and only HD~17092 and HD~102272 are already known as planet hosting stars (Niedzielski \etal\ \cite{N07}, \cite{N08}).

Observations were made with the Hobby-Eberly Telescope (HET) (Ramsey \etal\ \cite{R98}) equipped with the High Resolution Spectrograph (HRS) (Tull \cite{T98}) in the queue scheduled mode (Shetrone \cite{S07}). The spectrograph was used in the R=60.000 resolution mode and it was fed with a 2 arcsec fiber. The spectra consisted of 46 Echelle orders recorded on the 'blue' CCD chip (407.6-592 nm) and 24 orders on the 'red' one (602-783.8 nm). Typical signal to noise ratio was 200-250 per resolution element. The basic data reduction were performed using standard IRAF \footnote{IRAF is distributed by the National Optical Astronomy Observatories and operated by the Association of Universities for Research in Astronomy, Inc., under cooperative agreement with the National Science Foundation.} tasks and scripts. 

To estimate fundamental stellar parameters such as effective temperature, surface gravity, microturbulence velocity and metallicity we use a purely spectroscopic method  based on analysis of iron lines.  We therefore need equivalent widths (EW) of  hundreds of Fe~I and  Fe~II lines for every object. To this end we use DAOSPEC -- a code which measures EWs in an automatic manner (Pancino \& Stetson \cite{PS07}). We conclude that the EWs obtained with DAOSPEC are in good agreement with our previous hand-made measurements.

\section{Method and results}
The atmospheric parameters for program stars were obtained with the TGVIT algorithm developed by Takeda \etal\ in \cite{T02} and updated in \cite{T05a} and \cite{T05b}. This purely spectroscopic method is based on analysis of iron lines and relies on three conditions resulting form assumption of LTE that have to be satisfied: (a) the abundances derived from Fe~I lines may not show any dependence on the lower excitation potential (excitation equilibrium); (b) the averaged abundances from Fe~I and Fe~II lines must be equal (ionization equilibrium);  (c) the abundances derived from Fe~I lines may not show any dependence on the equivalent widths (matching the curve of growth shape). In our case these three requirements are fulfilled. 

The sample under consideration is composed of 22 red giants showing RV variations of 250 m s$^{-1}$ > RV > 40 m s$^{-1}$ (possible planetary-mass companions hosts) and a control sample of 58 giants with RV < 40 m s$^{-1}$, presumably single stars. 

For the sample of 80 stars we determined intrinsic  stellar parameters using $\approx$100 Fe~I and $\approx$10 Fe~II lines from the lists presented in Takeda \etal\ (electronic table) identified in the 'blue' region of HET/HRS spectra. Lines stronger than 150 m\AA{} at excitation potential less than 0.5 eV were rejected to avoid a disturbing of mean iron abundance trend relative to the excitation potential and EWs.

Metallicity was estimated with a satisfactory accuracy of $\sigma$[Fe/H] = 0.06 dex, on average,  for all objects. We obtained the mean values of [Fe/H] = -0.41 dex for the sample of stars exhibiting RV variations and [Fe/H] = -0.26 dex for the RV-stable giants from the control sample (or -0.25 dex if we take into account subgiants also, \ie\ stars on Fig.~1 with log(L/L$_{sun}$) < 1.3).

These results show the tendency suggested by Pasquini et al. (2007) but exaggerated. Giants demonstrating significant RV variations and thus suspected planet hosts are on average more metal-poor (Fig.~2) but the difference between the two samples is bigger than expected. As the difference in metallicity is much bigger that estimated metallicity uncertainty we conclude that the average metallicity of evolved stars with significant RV variations is 0.15 dex  (2.5 $\sigma$) lower than the mean value for the control sample of  the RV-stable stars from our sample.

\begin{figure}[htb]
\center
\includegraphics[width=7cm, angle=270]{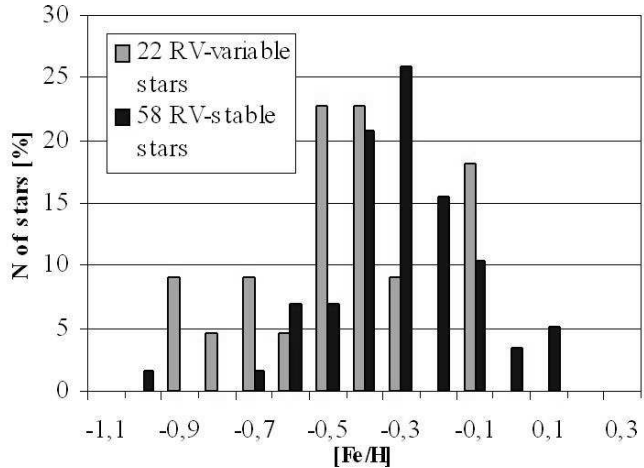}
\caption{Metallicity distribution for the sample of 80 giants. Stars with significant radial velocity variations (grey columns) are shifted in metallicity towards lower values comparing to RV stable objects (black columns).}
\end{figure}

The result is very preliminary. The future work (in progress) will extend the EWs measurements to the 'red' region of HET/HRS spectra increasing the number of identified iron lines and thus will verify the metallicity determination. The nature of the RV variations of the 22 possible planetary hosts has to be studied in detail as well before we confirm the metallicity depletion in planet hosting giants.

\section{Acknowledgments}
We thank Y. Takeda as well as P. Stetson and E. Pancino for making theirs codes available for us.
The Hobby-Eberly Telescope (HET) is a joint project of the University of Texas at Austin, the Pennsylvania State University, Stanford University, Ludwig-Maximilians-Universit\"at M\"unchen, and Georg-August-Universit\"at G\"ottingen. The HET is named in honor of its principal benefactors, William P. Hobby and Robert E. Eberly.

This project was  supported  by the Polish Ministry of Science and Higher Education grant 1P03D 007 30. PZ was supported by Polish Ministry of Science and Higher Education grant SPB 104/E-337/6.


\begin{thebibliography}{}
\bibitem[2005]{F05} Fisher, D. \& Valenti, J. 2005, ApJ, 622, 1102  	
\bibitem[1997]{G97} Gonzalez, G. 1997, MNRAS, 285, 403  	
\bibitem[2007]{H07} Hekker, S. \& Melendez, J. 2007, A\&A, 475, 1003  
\bibitem[2008]{H08} Hekker, S., Snellen, I.A.G., Aerts, C. \etal\ 2008, A\&A, 480, 215  
\bibitem[2007]{N07} Niedzielski, A., Konacki, M., Wolszczan, A. \etal\ 2007, ApJ, 669, 1354  
\bibitem[2008]{N08} Niedzielski, A., Gozdziewski, K., Wolszczan, A. \etal\ 2008, ApJL, submitted  
\bibitem[2007]{PS07} Pancino, E. \& Stetson, P.B. 2007, http://cadcwww.hia.nrc.ca/stetson/daospec  
\bibitem[2007]{P07} Pasquini, L., Dollinger, M.P., Weiss, A. \etal\ 2007, A\&A, 473, 979  
\bibitem[1998]{R98} Ramsey, L.W., Adams, M.T., Barnes, T.G. \etal\ 1998, Proc. SPIE, 3352, 34  
\bibitem[2001]{S01} Santos, N.C., Israelian, G. \& Mayor, M. 2001, A\&A, 373, 1019  
\bibitem[2004]{S04} Santos, N.C., Israelian, G. \& Mayor, M. 2004, A\&A, 415, 1153  
\bibitem[2007]{S07} Shetrone, M., Cornell, M., Fowler, J. \etal\ 2007, PASP, 119, 556  
\bibitem[2002]{T02} Takeda, Y., Ohkubo, M. \& Sadakane, K. 2002, PASJ, 54, 451  
\bibitem[2005a]{T05a} Takeda, Y., Ohkubo, M., Sato, B. \etal\ 2005a, PASJ, 57, 27  
\bibitem[2005b]{T05b} Takeda, Y., Sato, B., Kambe, E. \etal\ 2005b, PASJ, 57, 109  
\bibitem[1998]{T98} Tull, R.G. 1998, Proc. SPIE, 3355, 387
\end{thebibliography}
\end{document}